\documentclass[nonacm]{acmart}
\setcopyright{none}
\renewcommand\footnotetextcopyrightpermission[1]{}
\usepackage{lmodern}
\makeatletter                   
\def\mdseries@tt{m}             
\makeatother                    

\usepackage{listings}
\usepackage{mi}
\usepackage{preamble}
\usepackage{tikz}
\usepackage{stmaryrd}
\graphicspath{{./figs/}}

\startPage{1}

\setcopyright{none}

\usepackage{booktabs}   
\usepackage{subcaption} 

\begin{document}
\sloppy                         

\title[\textit{Managed Information}]{Managed Information: A New Abstraction Mechanism for Handling Information in Software-as-a-Service}

\author{David H. Lorenz}
\affiliation{
  \department{Department of Mathematics and Computer Science}              
  \institution{Open University of Israel}            
  \streetaddress{1~University Rd.,}
  \city{Ra'anana 4353701}
  \postcode{P.O. Box 808}
  \country{Israel}                    
}
\email{lorenz@openu.ac.il}          

\author{Boaz Rosenan}
\affiliation{
  \department{Dept. of Computer Science}             
  \institution{University of Haifa}           
  \streetaddress{199~Aba Khoushi Ave., Mount Carmel,}
  \city{Haifa 3498838}
  \country{Israel}                   
}
\email{brosenan@gmail.com}         
\begin{abstract}
Management of information is an important aspect of every application.
This includes, for example, protecting user data against breaches
(like the one reported in the news about 50 million Facebook
profiles being harvested for Cambridge Analytica), complying with
data protection laws and regulations (like EU's new \emph{General Data
Protection Regulation}), coping with large databases, and retaining
user data across software versions.  Today, every application needs to
cope with such concerns by itself and on its own.

In this paper we introduce \emph{Managed Information (\MI)}, an
abstraction mechanism for managing extra-functional data related concerns,
similar to how managed memory today abstracts away many memory related
concerns.  \MI{} limits the access applications have to user data, which,
in return, relieves them from responsibility over it.  This is achieved
by hosting them on a \emph{Managed Information Platform (\MIP)}, and
implementing their logic in a language that supports \MI{}.  As evidence
for the feasibility of \MI{} we describe the design and implementation of
such a platform.
For demonstration of \MI{},
we describe a simple social network application built with it. The
implementation is open source.

\end{abstract}


\maketitle
\fancyfoot[RO,LE]{OKAY}
\thispagestyle{empty}

\section{Introduction}\label{sec:intro}

Abstraction is a key concept in software. Being able to
abstract away aspects of software has been an enabler to 
many advancements in the field of \emph{Software Engineering} and to 
the construction of complex software systems~\cite{Jackson:2016:SAL}. 
Choosing the right level of abstraction presents a trade-off in designing 
the architecture of the software system. On the one hand, the higher
the level of abstraction is, the more aspects it will abstract away but 
the more rigid it becomes. On the other hand, the lower the level of 
abstraction is, the more control it provides over the end result
but the more effort it requires of its consumer.

As an example, an \emph{Instruction Set Architecture (ISA)} (e.g.,
Intel's x86 family) is commonly thought of as the lowest level of
abstraction available to software. It abstracts away concerns handled
by the hardware, such as register forwarding and memory address
translation. Meanwhile, it leaves many other aspects for programmers and
compilers consuming this abstraction to deal with on their own. For
example, this abstraction treats memory as a flat array of cells. It does
not define how memory is to be structured, and how allocations are made.

As another example, a virtual machine (e.g., the \emph{Java Virtual
Machine (JVM)}) often provides a higher level of abstraction, treating
memory as a graph of objects pointing to one another. This higher
level of abstraction allows memory to be managed (e.g., the \Java{} langauge with 
\emph{managed memory} and garbage collection).
By defining what objects are and how they reference each
other, the abstraction can take ownership over memory deallocation,
and provide better guarantees of memory {safety}, such as protection
against buffer overruns. This increased level of abstraction comes
at a price. Programs written over the JVM lose the ability to access
memory directly, and also lose fine-grained control over run-time,
as the garbage collector can kick in at any time.

Computer programs can also define abstractions. \emph{Content Management
Systems (CMSs)}, such as WordPress, Drupal, or MediaWiki, for example,
abstract away many of the common concerns related to building and
serving websites, while allowing their users to concentrate on the
contents they wish to present.
Like every abstraction, this too comes at the price of control. CMSs
typically allow content to be freely edited, but give very little
control over functionality. For example, users creating blogs over
WordPress can customize their blogs with predefined options programmed
into WordPress, but have no way of creating new social features
without breaking the abstraction level, implementing them using \PHP{}
code.

\subsection{Abstracting Information}

Our work is concerned with the level of abstraction available for
handing \emph{information} in software. By {information} we refer generally 
to data-related concerns that are extra-functional in nature but nonetheless addressed
individually by every application separately.
Some concerns relate to the amount of information applications need to
handle, and to the challenges in maintaining efficiency and availability
while handling large amounts of data. Other concerns relate to securing
information and to regulation regarding the integrity
and confidentiality of the information stored.

The lack of sufficient abstractions for information intensive software 
is especially evident in today's \emph{Software as a Service}~({\SaaS})
applications~\cite{Buxmann:2008:SAS,Turner:2003:TSI}.
%
%
In terms of information, \SaaS{} applications are ``unmanaged.''
Each application is responsible on its own for properly protecting user
data, and for proofing itself against known vulnerabilities. \SaaS{}
applications have to handle data denormalization and data migration.
And there are data protection regulations to which each application
needs to comply.

To mitigate this problem, we contribute in this paper an
abstraction mechanism named \emph{Managed Information}
(\MI). This abstraction allows for \SaaS{} applications to
be implemented, while the following five concerns regarding
the management of information are abstracted away:
\begin{itemize}
\item\emph{Access Control}
In \SaaS,
{access control} is coupled with {authorization}.
  \MI{}~provides a separation, where
  {authorization} is owned by the application, and {access control} is
  abstracted away.
\item\emph{Vulnerabilities}
Security
  {vulnerabilities} can be found anywhere in the software, and
  therefore, protecting against them involves all developers.
  \MI{}~makes {vulnerabilities} in the application code irrelevant 
  by guaranteeing that the information remains secure
  even if the application itself acts on behalf of an attacker.
\item\emph{Denormalization} 
To handle large amounts of information,
  \SaaS{} application developers are often required to {denormalize} their data,
  by maintaining redundant representations of it. This redundancy adds
  complexity to the implementation, and is prone to bugs.
  \MI{}~allows
  for {denormalization} to be abstracted away. The application only
  defines the relations that need to be maintained.
\item\emph{Software Evolution}
There is a tension between the desire for
  the application to {evolve}, and the need to retain old data, along
  with assumptions made on it.
  With~\MI{} the software may {evolve} as its developers desire, while
  information is retained with its integrity and confidentiality
  assumptions unchanged.
\item\emph{Regulatory Compliance}
Regulators are becoming more and more
  involved in management of information. GDPR~\cite{GDPR:2016} is one example of
  {regulation} that states what rights users must be given over
  information they contribute.
  \MI{}~provides a way to implement
  features required for {regulatory compliance} in an
  application agnostic manner. This allows applications to comply with
  such regulations without taking any specific actions.
\end{itemize}

\subsection{Conceptual Contribution}
The main research question addressed in this work is one of feasibility~\cite{Shaw:2002:WMG}: \textit{is it possible
at all to accomplish a level of abstraction that supports~\MI{}?}
If so, to what extent these five concerns or others could be
managed in~\MI{}? To that end, we present a conceptual framework for~\MI. To
validate the concept of~\MI{} we implemented a fully functional, open
source, proof-of-concept \MI{} platform for \SaaS{} applications.
With respect to the 5
concerns listed above, our proof-of-concept implementation validates
the feasibility of~\MI{}. We also provide a demonstration of how \MI{}
works on  \TweetMI{}, a Twitter-like example, which is simple yet
representative of complex social network \SaaS{} applications in reality.

\section{Conceptual Framework}\label{sec:nutshell}

In the knowledge
hierarchy~\cite{Rowley:2007:TWH},
	also known as the Data/Information/Knowledge/Wisdom
	(DIKW) hierarchy,
\emph{information} resides directly on top of the
foundation of \emph{data}.  Data refers to raw
observations, whereas information refers to meaning
logically inferred from data or from processed data.
In these terms, \emph{Managed Information} (\emph{\MI})
abstracts over {information}, i.e., the {meaning}---not
just the {form}---of data.

From a programming perspective, \MI{} relieves
the application developer of most of the responsibility
over user data.  Through the use of abstraction, \MI{} lets the
application programmer decouple the application logic from
the data, similar to how a garbage collector frees the
programmer from most of the responsibility over memory
management.  Indeed, with \MI{} eight out of the nine
application-level vulnerabilities listed in the 2017's
OWASP Top 10 \cite{OWASP:2017:T10} would be abstracted
away by the language implementation
(\ref{sec:disc-vulnerabilities}).

From a data management  perspective, \MI{} is an
abstraction mechanism that subsumes and supersedes
that of a \emph{Database Management System (DBMS)}.
In traditional applications, a DBMS is often used as the
primary abstraction over the state of the application,
i.e., its \emph{data}. While a DBMS abstracts away the
nuts and bolts of how data is stored and retrieved, it
manages only the data's bits and structure and not its
implied \emph{information}.

We explain the \MI{} abstraction in three steps
(corresponding to \ref{sec:ap,sec:mip,sec:mi}).

\subsection{From Data to Information}\label{sec:ap}

The first step allows for denormalization and data
migration to be decoupled from the application logic.
This is achieved by lifting the repository from handling
data to handling information by associating the data
with logic capable of answering questions regarding it.
An \emph{Application Platform}~(\AP) capable of answering
questions regarding the data becomes a repository for
information:
 \begin{definition}[\AP]\label[def]{def:AP}
  An \emph{Application Platform} is software that provides
  the following operations: \begin{enumerate} \item Create,
  Read, Update and Delete (CRUD) data.  \item Create,
  Update and Delete logic, which defines how questions and
    data translate into answers.
  \item Perform queries: given a question, provide an
  answer based on
    the logic and the data.
  \end{enumerate}
 \end{definition}

Similarly to a DBMS, an \AP{} supports storage and
retrieval of data. However, unlike a DBMS, an \AP{}
also stores logic which effectively gives the data
meaning. A DBMS, in contrast, stores the data without any
interpretation, i.e., the queries answered by a DBMS
refer to the data symbolically (e.g., fields in
tables) rather than to the meaning it represents.

\subsection{From Unmanaged to Managed Information}\label{sec:mip}

The second step allows for access control to be decoupled
from the application logic. This is achieved by adding
an ownership model to the \emph{Managed Information
Platform}~(\emph{\MIP{}}), giving each piece of data an
owner and an intended audience:
 \begin{definition}[\MIP]\label[def]{def:MIP}
  A \emph{Managed Information Platform} is an \AP{}
  that meets the following additional requirements:
  \begin{enumerate}\addtocounter{enumi}{3}
  \item\label[req]{req:integrity} Data is attributed to
  users. Any user can perform any CRUD
    operation on data, as long as the data is attributed
    to that user.
  \item\label[req]{req:trust} Users and logic can specify
  whose data and logic to
    trust. Untrusted data and logic is not taken into
    consideration.
  \item\label[req]{req:confidential} Users who create data
  can specify who can read it. Queries made
    by a user only consider data that user is allowed
    to read.
  \end{enumerate}
 \end{definition}

In \Ref{def:MIP} a \emph{\MIP{}} is
an \AP{} that also addresses \emph{data
integrity}~(\ref{req:integrity,req:trust}) and \emph{data
confidentiality}~(\ref{req:confidential}).  Data stated by
a user is always accompanied by (at least) two pieces of
information: {\ONE}~an attribution to the user who created
the data, and {\TWO}~the specification of who can read this
data. Mathematically, both can be represented as \emph{sets
of users}. We therefore call the attribution of a piece of
data (fact)---its \emph{writers-set}, and the specification
of who is allowed to read it---its \emph{readers-set}.
We will refer to a single, fact (self-contained piece
of data) that has a writers-set and a readers-set as an
\emph{axiom}.

By defining the writers-set as a set, rather than fixing it
to the identity of a single user, we allow users to state
facts in the name of groups. For example, academics can
state facts in their own names, or in the names of their
departments or universities. If they state something in
their own names, no one else can update or remove their
statement. However, if they use a larger group, other
members of that group can modify or remove that fact in
case they do not agree with it.

\subsection{Managed Information}\label{sec:mi}

We can now already formulate what we mean by \MI{}
in the context of \SaaS{}:
 \begin{definition}[\MI]\label[def]{def:MI}
  \emph{Managed Information} is a software architecture
  and programming model for \SaaS{} applications, in which
  the server-side consists of a \MIP{}, which by itself
  forms a \emph{Platform-as-a-Service} ({\PaaS}).
 \end{definition}

\section{Approach}\label{sec:approach}
A key principle in our approach to enabling \MI{} is to distinguish
between what is subjective and what is objective
\emph{w.r.t.}~information.
  	
\subsection{Objectivity of Data}\label{sec:objectivity-of-data}

\SaaS{} applications that use multiple instances of ACID 
databases, or a single instance of a non-ACID database, 
do not have a well defined global state.
One approach that gains traction in recent years to
overcome the lack of well-defined state is \emph{event
sourcing}~\cite{Hohpe:2006:PWC,Fowler:2005:ES}.  Event
sourcing is an approach to the management of state in
applications, where the application's single source of
truth is the collection of \emph{change events}, and
partial state is generated on demand from these events.

The core philosophical principle behind event sourcing
can be abbreviated as: \emph{while state is subjective
(depends on the observer), change events are objective}.
To demonstrate this benefit of event sourcing, let us
consider a simple example. The state of an application
contains variable~$A$, which can be assigned different
values at different times. Now assume one user sets
variable~$A$ to~$1$ at about the same time another user
sets variable~$A$ to~$2$. In the traditional notion of
state, variable~$A$ can hold either~$1$ or~$2$, depending
on factors such as the exact timing, the way state is
updated, and the identity of the observer. However, the
two change events, $A\leftarrow 1$ and~$A\leftarrow 2$,
are well defined. When someone wishes to know the value
of~$A$, they can play the two events and decide, based on
their own perspective, what the value should be.

\MI{} makes state even more subjective, 
by annotating data with confidentiality requirements
(readers-sets, recall \ref{req:confidential} in \ref{def:MIP}).
This means that when viewed at the same
time from the eyes of different users, the same application
can seem to have multiple different states, since different
users are allowed to see different facts.
For example, consider the two change events $A\leftarrow 1$ and
$A\leftarrow 2$ from before, but this time, the former's readers-set
is the set $\left\{u_0\right\}$ for some user $u_0$, and for the
latter, the readers-set is the set $\mathbb{U}\setminus\left\{u_0\right\}$, where
$\mathbb{U}$ represents the universal set of users. The state that
arises from the accumulation of both events is subjective, and depends
on the observer. User $u_0$ will observe $A$ to have the value $1$,
while any other user will observe $A$ as $2$.

To cope with the subjective nature of state, a \MIP{} should only
store events and not accumulate them to state. This means that all the
processing done inside a \MIP{} should be based on events. Unlike the
\MIP{} itself, clients can receive events readable by the user they
represent, and accumulate them to form state, as seen by that user.

\subsection{Objectivity of Logic}\label{sec:objectivity-of-logic}

In \MI, not only data is subjective, but also the logic that gives it
meaning. For example, social networks often
provide their users with \emph{timelines} or \emph{feeds}, containing
information coming from their social environment. The logic for
creating this timeline or feed is specific to each social network, and
often changes with time. Therefore, even if all facts contributing to
a user's timeline are known, the contents of a user's timeline is not
objective, but rather dependent upon the exact version of the logic
that generates it.

This notion becomes extremely important when dealing with data
denormalization. To shorten the time it takes a social network to
present a timeline to a user, many social networks prepare their
timelines ahead of time~\cite{Evans:2010:CBE}, populating them with social data such as
tweets and statuses as they come in. To do so,
they need to apply logic.

However, applications evolve with time. The developers of a social
network may change the logic for constructing a timeline as time goes
by. Therefore, the version of the software that was in place when
some tweet~$T_1$ was written (and placed into the denormalized
timeline) may not be the same as the software version at 
the time when some tweet~$T_2$ was written, and placed into the same
timeline, and that may not be the same version as the one at
the time in which the user has queried his or her
timeline.

To overcome this problem, application developers often develop data
migration routines that are applied to pre-existing data during a
migration from one version to another, modifying the form of the
derived data (e.g., the pre-calculated timeline) according to the new
logic. This makes sure a query will always reflect the version at the
time it was made.

Unfortunately, this kind of migration process comes at a high
price. The migration routine is a computer program intended to run
only once, but still it needs to be written and tested thoroughly to
make sure its output is consistent with what the new logic would have
produced.

Another challenge is to perform this migration without stopping the
application, with new data coming in and queries being made while the
migration process is running.

By storing logic alongside data, \MI{} makes both denormalization and
migration, managed aspects. However, to do so properly, it has to have
an objective notion of logic.

Our approach for making logic objective involves two steps: ({1})~using
content-addressable code, and ({2})~restricting logic to being
purely-declarative.

\subsubsection{Content-Addressable Code}
Content-addressable storage~\cite{Merkle:1988:DSB} is a practice by
which data objects are stored using keys derived from applying a
cryptographic hash function on their content. If a data object
references another, it does so by mentioning its hash, and therefore
this hash is considered when calculating the hash of the referencing
object.

Content addressing can be used with code as well as data. Consider a
module system, in which the name of a module is derived from a
cryptographic hash of the code it contains. When such a module
references another module, the hash of the referenced module is
written explicitly in the referencing module, making it part of its
content, affecting its hash.

Content-addressable code helps make logic objective. While the term
\emph{timeline} may be subjective, depending on a specific version of
a specific social network, the timeline associated with a specific
hash~$h_1$ is objective, since $h_1$ defines a specific module that
defines a specific version of the logic building the timeline, along
with all its dependencies.

When a new version comes along, it does not change the logic
associated with~$h_1$, but rather creates new logic, associated with a
new hash~$h_2$. If we name the derived data artifacts (e.g., the
pre-calculated timeline) after the logic that created it, the two
versions of the logic would create two distinct artifacts, one
prefixed with~$h_1$, and the other, with~$h_2$. Being distinct, they
can coexist.

This simplifies the migration process. First, migration to the second
version is as simple as applying the new logic to all existing
data. Second, the old derived data can be used while migration is
taking place.

\subsubsection{Purely-Declarative Logic}
Another important aspect of keeping logic objective is restricting it
to being purely-declarative. Logic can be applied multiple times to a single
piece of data. For example, the logic of placing a tweet in a timeline
can be applied when the tweet is being written, then (with some
modification) during migration to a new version, and finally, when the
tweet is removed. To make sure all applications of the logic yield
consistent results, we need to make sure the logic itself is not
affected by the outside world in any way.

If, for example, the logic for calculating the timeline considered the
time of processing to populate the timeline, the time the migration
process considered a certain tweet would affect the way it is
presented in timelines.

\subsection{Application Architecture}\label{sec:architecture}
%
 \begin{figure} \centering
	\subcaptionbox{Traditional
	architecture\label{fig:traditional:architecture}}
	{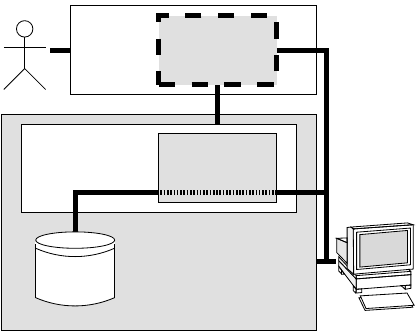\hspace{0.01\columnwidth}}
	\subcaptionbox{\MI{}
	architecture\label{fig:mi:architecture}}
	{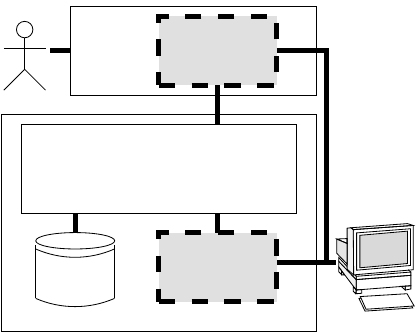\hspace*{\fill}}
	\caption{Client-server architecture comparison.
	A dashed frame indicates sandboxed code.
	A gray background indicates responsibilities of
	the application code.  The numbers (1) and (2)
	denote interfaces.}\label{fig:architecture}
 \end{figure}
Similar to a traditional client/server application
architecture (\ref{fig:traditional:architecture}), the
\MI{} application architecture (\ref{fig:mi:architecture})
has an application-specific (depicted in gray) client-side,
running on an application-agnostic (depicted in white)
client. This could be \JavaScript{} code running on
a browser or a mobile client app running on a mobile
OS. Having the application's client-side as payload over a
client platform has advantages in protecting user privacy,
as it allows executing it from within a sandbox (depicted
as a dashed frame)---a controlled environment which limits
its access.

However, \MI{} differs from the traditional client/server
architecture in the role the application logic plays on the server
side.
In a traditional applications, the
server-side is at the complete control of the application
developers, and under their complete responsibility. They
may use off-the-shelf software, such as web frameworks or
databases to simplify some aspects of the application, but
these are used, misused, or bypassed at their discretion.
While web frameworks can play an important role in keeping
the application code focused on application-specific
concerns (e.g., by handling many common concerns within
the framework), they do not limit the application code
in any way. Once invoked, the application logic is in
control, typically having the same level of access as the
framework itself. This level of access makes application
logic subject to security vulnerabilities and data leaks.

In contrast, the \MI{} architecture stores the
application's server-side logic as payload on a \MIP,
similarly to the way the client-side logic is stored on
the client. Just like the client, it gives the \MIP{} the
opportunity to run it from within a sandbox.  The sandboxed
server-side logic does not have direct access to the data,
nor to the client. All communications are brokered by
the \MIP{}'s logic, labeled \emph{Information Manager}
in \ref{fig:mi:architecture}. The \IM{} allows clients to
manipulate user data (subject to the restrictions posed in
\ref{req:integrity}), and answer client queries (subject
to the restrictions posed in \ref{req:confidential}). To
be able to answer these queries, the \IM{} consults the
application logic, providing it all the information it
needs. The answers provided by the logic are used by the
\IM{} to either answer a query directly, or store answers
for a later date.

Much of the stake regarding the feasibility of \MI{} lies
on the question of what kind of restrictions we would
like to pose on an application's server-side logic. As
in the case of \JavaScript{} on a browser, we need these
restrictions to be, on the one hand, restrictive enough
to prevent the logic from being a security concern, and
on the other hand, permissive enough to allow the logic
of any application to be expressed.

The integrity of the application is its own responsibility
(according to \ref{req:trust}).  However, confidentiality,
i.e., the prevention of data leakage, needs to be handled
by the sandbox. We distinguish between two kinds of leaks:
\emph{direct} and \emph{indirect}.  Direct leakage means
that when processing the data it wishes to leak, the code
transmits it on some channel to a party ready to receive
it. Indirect leakage is done in two steps. At the first
step, when processing one user's data the code stores the
data it wishes to leak in some temporary storage, and then,
when processing another user's data (say, the attacker's),
it fetches the stored data and returns it as the answer.

\MI{} prevents both kinds of leakage by disallowing side
effects. Side effects are needed to either transmit data
or to store it aside. Specifically, \MI{} restricts the
application's server-side logic to a purely-declarative
programming language thereby preventing data leakage,
while still allowing applications to process user data.

\section{Design}\label{sec:design}

As a proof-of-concept for \MI{} we have implemented \Ax{},
a \MIP{} intended to be provided as a service~(\PaaS). In this section,
we describe key characteristics of the design of \Ax{} intended to make
our approach to \MI{} practical and scalable.

\subsection{A DSL for Defining Logic}
To allow applications to define their logic, \Ax{}
provides a purely-declarative, logic programming
DSL,\IGNORENOTE{%
	The definition of the DSL is left outside the scope of
	this paper due to lack of space. {For the more technically
	inclined reader, discussion of the syntax and semantics
	can be found in the documentation, e.g.,
	\SANITIZEDURLNOTE{}{http://axiom-clj.org/cloudlog.html}}.
}
implemented as an internal DSL over
\Clojure.  With this DSL, application logic is defined
as a set of rules. A rule is a declarative definition of
the relationship between facts it takes as input, and the
derived facts it produces as output.

For example, consider a simple social network in which users can post
``tweets," and can follow other users, such that the tweets made by
the users they follow would appear in their timelines.
The logic of such a network can be defined using a rule that matches
facts corresponding to the phrease ``$u_1$ follows $u_2$'', and facts
corresponding to ``$u_2$ tweeted $T$'',
and generate a timeline entry corresponding to ``$T$ should appear in
$u_1$'s timeline''.
This timeline entry is a derived fact named after the rule.

The DSL rules also permit the use of \emph{guards},
which allow filtering of
facts based on their values, the derivation of new values, and
extraction of multiple values from a single value (e.g., indexing
tweets according to hash-tags mentioned in them).

One specific guard provided in our DSL (the \lstinline$by$ guard)
is responsible for integrity. It asserts that a fact the rule relates
to is attributed to a certain user. For example, it can be used to
assert that the fact ``Alice follows Bob'' is indeed attributed to
Alice. Facts that do not meet the criterion posed by the
\lstinline$by$ guard are ignored by such a rule.

\subsection{Permacode}\label{sec:permacode}

To allow logic to be objective,
we introduce a novel module system named
\emph{\Permacode}{\SANITIZEDURLNOTE{.}{http://axiom-clj.org/permacode.html}}
	In \Ax{} 
	\Permacode{} (short for ``permanent code'') module definitions are
	guaranteed to always behave the same way.
		\Permacode{} derives its name for the term ``permalink'', which refers
		to a URL that is guaranteed to always link to the same content.
With \Permacode, expressions consisting of nothing but
constants and definitions coming from \Permacode{} modules
will always evaluate to the same values.

This is achieved using a combination of the two
elements described in \ref{sec:objectivity-of-logic}.
To achieve purity, \Permacode{} provides a \DSL{} that is
a purely-functional subset of \Clojure. It presents one
macro (\lstinline$pure$), which examines the underlying
code at compile time and validates it against a white-list
of pure special forms and standard-library functions. If valid,
the underlying code is returned unchanged.

To achieve content addressing, \Permacode{} uses
a cryptographic hash function (SHA-256) on the contents of a module to
determine its name.
One
\Permacode{} module can reference another, in which case
the hash code of the referenced module becomes part of
content of the referencing module, and thus affects its
own hash code. This makes \Permacode{} modules form a
\emph{Merkle Tree}~\cite{Merkle:1988:DSB}.

\Ax{} takes application logic
in the form of \Permacode{} modules. By doing so, it
guarantees the logic to lack any side effects, and to
never change. By restricting the definition of the DSL to the
declarative subset defined by \Permacode, \Ax{} is able to accept
rules and can guarantee that logic that runs from within
its guards will not harm confidentiality.

Because \Ax{} names the derived facts after the rules
that created them, their names change as their code
is updated. As result, derived facts from different
versions of the application can coexist, and updates to one
kind of derived facts are always made according
to the same logic.

This also has integrity benefits. In case multiple applications are
hosted on the same instance of \Ax{}, it is possible one will try to
create fake results, and masquerade them as legitimate results
provided by another application. However, by using \Permacode{} this
becomes practically impossible, since
forging a derived fact becomes
as hard as performing a second pre-image attack on
SHA256~\cite{Rogaway:2004:CHF,Gilbert:2003:SAS}.

\subsection{Event Sourcing}\label{sec:event-sourcing}

For the sake of keeping data stored in \Ax{} objective (as discussed
in \ref{sec:objectivity-of-data}),
we designed \Ax{} as an event
sourcing system. An \Ax{} event corresponds to the concept
of an axiom, as described in \ref{sec:mip}. We define an
event as a tuple $\left\langle d,w,r,c,t\right\rangle\in
E$, where~$d$ is a data tuple representing either a
fact or a rule, $w\subseteq\mathbb{U}$ is a writers-set,
$r\subseteq\mathbb{U}$ is a readers-set, $c\in\mathbb{Z}$
represents the quantitative change this event represents
($c=1$ for addition, $c=-1$ for removal), and~$t$ is a
unique identifier.

\Ax{} takes the collection~$E$ of these events as its
source of truth. Partial state, as seen by user~$u$,
can be built in two steps. First, we define the
subset~$V_u\subseteq E$ of the events visible by user~$u$
as:
$V_u=\left\{\left\langle d,w,r,c,t\right\rangle\mid
\left\langle d,w,r,c,t\right\rangle\in E \wedge u\in
r\right\}$.  Then we translate these events to partial
state by grouping~$V_u$ according to~$d$, $w$, and~$r$,
summing~$c$.

\Ax{} itself never accumulates state. It receives events from the
client, and sends events back. \Ax{} provides a client-side
library{\SANITIZEDURLNOTE{}{http://axiom-clj.org/axiom-cljs.html}} which
provides abstractions for accumulating events into state, as seen from
the eyes of a single user, on the client side.

\subsection{Gateway Tier}\label{sec:gateway-tier}

For sending and receiving events, the
clients communicate with \Ax{}'s \emph{gateway
tier}{\SANITIZEDURLNOTE{.}{http://axiom-clj.org/gateway.html}}
The gateway tier can be seen as the boundary
between \Ax{} and the outside world. As such,
it handles the connection to the clients (through
WebSockets~\cite{Qveflander:2010:PRT}), authentication,
which is abstracted out through a plug-in architecture,
and access control.

\Ax{}'s access control is decoupled from any application,
and follows \ref{req:integrity,req:confidential} of
\ref{def:MIP}. This means that when a client which
authenticated as user~$u$ sends event $\left\langle
d,w,r,c,t\right\rangle\in E$ to \Ax{}, the gateway
tier will accept it if and only if~$u\in w$. If a
client authenticated as user~$u$ registers to receive a
subset~$E'\subseteq E$ of the events stored by \Ax{},
the gateway tier will send the client only those~$u$
is allowed to see, i.e., $E'\cap V_u$.

To implement access control, \Ax{} takes a two-step approach. First
it defines \emph{named groups}. These are conceptual groups of users
which are named after a common property. For example, all ``friends'' of
user~$u$ can be placed in a named group.

Named groups are populated using rules, and are
named accordingly, prefixed with the hash of their version.
This makes their content objective, e.g., defines the exact
criteria for user~$u_1$ to be ``friends'' with user~$u$ in a specific
version of the application.

The second step is constructing a representation of set~$U$ associated
with a certain user. This set is an intersection of the singleton set
consisting of the user (represented by his or her user ID), and all
the named-groups this user is a member of. This set represents the
smallest set the user is a member of, in the sense that for every
set~$U'$, $u\in U'\leftrightarrow U\subseteq U'$.

\Ax{} uses
\emph{intersets}{\SANITIZEDURLNOTE{}{http://axiom-clj.org/cloudlog.interset.html}}
to represent sets. In this representation, named groups are
represented symbolically (i.e., mention named groups, and do not list
the actual users included in them).
Intersection of two intersets, and
testing if one interset is a subset of another are implemented
efficiently on intersets.

\subsection{Information Tier}\label{sec:information-tier}

When an event coming from the client passes
through the gateway-tier, it reaches \Ax{}'s
\emph{information-tier}. This tier takes its name from
the fact that it holds both data and logic. At the center
of the information-tier, \Ax{} uses an event broker
(RabbitMQ){\URLNOTE{}{https://www.rabbitmq.com/}}
that spreads events through its different
components. Events are stored in a NoSQL database (Amazon's
DynamoDB){\URLNOTE{,}{https://aws.amazon.com/dynamodb/}}
but are also processed to produce new events.

\subsubsection{Migration}

New versions of application logic are provided by events
pointing \Ax{} to a git repository containing the new
version. This starts a migration process, in which
\Ax{} fetches the sources from the git repository,
publishes each source file as a \Permacode{} module,
and
applies the logic on all existing fact
events{\SANITIZEDURLNOTE{.}{http://axiom-clj.org/migrator.html}}

\subsubsection{Topologies}

Once the migration is complete, \Ax{} deploys a topology
to handle further events coming in real time. \Ax{} uses
Apache Storm~\cite{Iqbal:2015:BDA}, a platform designed
for executing real-time analytics on big data. Although not
intended for this use, \Ax{} uses it to apply application
logic at real time. Storm is known to perform computation
with low latency and high throughput, be fault tolerant
and highly available. By using it, \Ax{} inherits these
guarantees.

\subsection{Event Processing}\label{sec:event-processing}

Regardless of when events are being processed,
be it during a migration of new logic, or when
an existing topology processes a new event,
\Ax{} uses the same mechanism for processing
events{\SANITIZEDURLNOTE{.}{http://axiom-clj.org/cloudlog-events.html}}

Given rule~$R$ we define its
initial event as $e_R=\left\langle
R,\textit{ns}(R),\mathbb{U},1,t\right\rangle$, where
$\textit{ns}(R)$ is the namespace~$R$ is defined in
(recall that~$R$ is defined in a \Permacode{} module
for which the name is based on a hash of the module's
content), $\mathbb{U}$ is the universal user set, and~$t$
is an arbitrary unique ID.

\subsubsection{Product}\label{sec:product}

Given a fact event $e_F=\left\langle
F,w_F,r_F,c_F,t_F\right\rangle$ and a rule
event~$e_R=\left\langle R,w_R,r_R,c_R,t_R\right\rangle$ we
define the \emph{product} of these events $e=\left\langle
d,w,r,c,t\right\rangle=e_R\otimes e_F$ as follows: $d$
is the result of applying~$R$ to~$F$. This can be either
a representation of a residual rule (in case of a join),
or a derived fact.  Because the rule is responsible for its own integrity,
the resulting
event's writers set $w=w_R$ is taken directly from the
rule, and the fact's writers set is ignored.

For the readers set, we take an intersection: $r=r_R\cap
r_F$. A user~$u$ may read~$d$ if and only if that user
is eligible to read both~$R$ and~$F$. With access to~$R$
and~$F$, user~$u$ could compute $d$. By setting~$r$ this
way we make sure the product operation does not change
the confidentiality properties of~$R$ and~$F$.

The quantitative change $c=c_R\cdot c_F$ is the numeric
product of the changes~$c_R$ and~$c_F$.  If the same
fact~$F$ has been added~$m$ times, and the same matching
rule~$R$ has been added~$n$ times, we get the result~$d$ as
if we applied every rule to every fact, $m\cdot n$ times.

The unique ID~$t$ is calculated as a hash of both IDs:
$t=\textit{hash}\left(t_R,t_F\right)$. This is to make sure
that the new event has a unique ID, but if the same product
is calculated twice, we will receive an event with the same
ID. This is important for fault tolerance, since Apache
Storm provides an \emph{at-least-once} guarantee, meaning
they may re-emit events to cope with failures. Having a
consistent event ID helps in making sure we do not count
duplicate events.

Note that we simplified the definition
	of the product for readability. In reality,
	applying a rule to a fact results in a set
	of results, and not (necessarily) a single
	result~$d$. This means that the product operation
	results in a set of events, and not necessarily
	a single event~$e$.

\subsubsection{Event Matching}\label{sec:event-matching}

An \Ax{} \emph{event processor} is an object that receives
a sequence of fact and rule events, and emits a sequence
of their products.  An event processor stores all events
it received, and accesses this storage as needed. When
receiving a fact event, it fetches all rule events
that share the same key, and computes their mutual
products. Similarly, when receiving a rule event, it
fetches all fact events with a matching key, and emits
all their products.

\Ax{} uses DynamoDB for storage. This database is
\emph{eventually consistent}, meaning an event that
has been stored at time~$t_1$ may not necessarily be
visible when queried at time~$t_2>t_1$, in cases when
$t_2-t_1<\Delta t$, where~$\Delta t$ is the time it takes
data to spread across all replicas inside DynamoDB. To
overcome this problem \Ax{}'s event processors use a
cache that stores events in memory for a given period
of time. \Ax{} configures its Storm topologies in such a
way that guarantees that events with a certain key will
always be processed by the same event processor, making
sure that even events that are emitted within~$\Delta t$
of one another will take each other into consideration.

\section{Evaluation and Discussion}\label{sec:evaluation}
For a concrete demonstration of~\MI{} we have implemented a simple
micro-blogging application inspired by Twitter as our
test application.  \TweetMI,
pronounced \emph{Tweet-Me}, is
a web application that allows its users to post tweets,
follow other users, and view their \emph{timeline}.
A user's timeline consists of: (1)~tweets made by the
user, (2)~tweets made by the user's followees (user whom
that user follows), and (3)~tweets in which the user's
\emph{handle} (\lstinline$@user-id$) appears.

In this section we reflect on the concerns described in
\ref{sec:intro} and discuss how they are met by~\MI{}
on the implementation of \TweetMI{} (\ref{appendix:tweetmi}).

\subsection{%
Responsibility Over User Data
}\label{sec:disc-access-control}

\TweetMI{}, the example application we chose for this
paper, is a social network. As such, its business logic
focuses on spreading information among users. Keeping
information private, however, may seem as going against the
very nature of social media. For this reason, \MI{}~makes
this the concern of the~\MIP{} and not the application.

This can be demonstrated in \TweetMI{}, by adding a
requirement for supporting restricted tweets -- tweets
visible only by users whom the tweet's author follows. For
example, if user \lstinline$"alice"$ posts a restricted
tweet, only the users \lstinline$"alice"$ follows can
see this tweet. This way, by following or un-following
other users, she can decide who is allowed to read her
restricted tweets.

Traditionally, adding such a feature would include changes
in the client (to introduce the ability to mark a tweet as
restricted), the data model (addition of a Boolean field
to indicate whether the tweet is restricted or not) and
the access control logic (avoid displaying restricted
tweets to users whom the tweet's author does not follow.

With~\MI, this change is made on the client-side
only. When creating facts, the client determines their
associated readers-sets. To allow restricted tweets to be
created, the client code must be updated to allow its
users the choice between setting the tweet's readers-set
to $\mathbb{U}$ for unrestricted tweets, and a named group
of users whom the user operating the client is following,
for restricted tweets.

%

By setting the readers-set according to the user's wishes
the application concluded it job in protecting this
tweet. Any further processing done with this fact is
guaranteed to respect this choice, by following the logic
described in \ref{sec:event-processing}.
This decoupling makes the application responsible for
\emph{authorization}, but not for access control.

\subsection{Protecting Against {Vulnerabilities}
}\label{sec:disc-vulnerabilities}

Not trusting the application with user data is key
in making~\MI{} secure. As an example, consider the
following vulnerability in \TweetMI{}. Suppose that
in one of the rules we accidentally try to call
\lstinline$eval$
(the \Clojure{} function that executes an s-expression)
on data taken from a fact to whcih the rule is applied.
Traditionally, this may leave the application
vulnerable to code injection attacks, allowing attackers
to execute their own code from within the rule. This would
enable an attacker to compromise confidentiality (leak
information out), integrity (make unauthorized updates),
and availability (cause a server failure).

Fortunately, \Ax{}, uses \Permacode{} as a sandbox around
application logic.  \Clojure{}'s \lstinline$eval$ function,
being imperative, is not allowed there. A replacement
function provided by \Permacode{} for evaluating \Permacode{}
functions
can be used from within rules, but only allows
purely-declarative code to run. Such code, even if injected
by an attacker, can only change the output of this rule,
thus impairing the integrity of the application, in a way
that is consistent with it having a bug. Confidentiality
and availability are not harmed.

As result, out of the \emph{OWASP
  Top~10}~\cite{OWASP:2017:T10},
the only vulnerability
that needs to concern an~\MI{} application is
\emph{A7:2017--Cross-Site Scripting}, which is the only
client-side vulnerability on that list. The rest need to
concern the~\MIP{} developers, but not the application
developers.

\subsection{%
{Denormalization}
}\label{sec:disc-denormalization}

\MI{} holds the key to freeing application developers from
having to worry about denormalization in that the~\MIP{}
encapsulates both the data and the logic the application
applies to the data. With access to both logic and data,
the~\MIP{} can be responsible for applying the logic,
whenever it needs to be applied, to the data.

In \TweetMI{}, three rules perform denormalization for
three different reasons. One rule takes facts representing
the phrase ``$u_1$ follows $u_2$'', and generates a
derived fact representing the phrase ``$u_2$ is followed
by $u_1$''. The rationale behind such a rule is that while
the raw fact is keyed by the follower ($u_1$), the derived
fact created by the rule is keyed by the followee
($u_2$). This facilitates answering the question ``who
follows user $u$?'' efficiently.

Another rule takes facts that correspond to the phrases:
``$u_1$ follows $u_2$'', and ``$u_2$ tweeted $T$'', and
creates derived facts corresponding to the phrase ``$T$
should appear in $u_1$'s timeline''.

This rule performs a join operation between the two kinds
of raw facts it takes. By performing it at
update-time, the application's query latency is improved
significantly.

The third rule takes facts that correspond to the phrase
``$u$ tweeted $T$'', and creates derived facts
corresponding to the phrase ``$u_i$ was mentioned in tweet
$T$''. The rule extracts the identity of zero or more
users from the text itself, by tokenizing it and
identifying tokens of the form \lstinline$@user-id$. This
allows adding tweets to the timelines of users who were
mentioned in them, without having to scan all the tweets
for such mentions.

In all three examples, the rules specified the format of
the denormalized data and the logic of how it is to be
derived from raw data.

Under the hood, these rules are interpreted as many small
procedures that operate at different times. For example,
when a tweet is being \emph{updated} (e.g., the text is being
edited), \Ax{} fetches all followers of the tweet's author
and updates the corresponding
derived facts. The rule tracking users mentioned in tweets will
create facts for new user handles, remove facts for handles
that were removed, and not touch ones that remain in the
updated text.

All these imperative procedures are derived automatically
by \Ax{}, without the rule having (or being able) to
specify them.

\subsection{Supporting Software {Evolution}
}\label{sec:disc-evolution}

Consider we wish to make two changes to \TweetMI{}:
(1)~refine the tokenization we perform on tweets when
searching for mentioned user handles,
and (2)~remove support
for restricted tweets (a feature we have added in
\ref{sec:disc-access-control}). 
These changes raise two challenges,
namely performing zero-downtime data migration, and
respecting authorization rules set by older versions,
while not complicating the logic in newer versions.

Performing these two changes to an application based on
\Ax{} is as simple as performing the changes in the source
code and re-deploying the application. The first change
is done in a rule. Since the code is located in
\Permacode{} modules, by updating the rule, its namespace,
which is derived from a hash of its content, changes. This
means that we did not update a rule, but rather created
a new one. After deploying the code, \Ax{}'s migrator
(\ref{sec:information-tier}) starts applying the rule
to all relevant facts, creating derived facts in the new
rule's namespace. These facts do not replace the old ones,
created by the old version of the rule. Both versions
co-exist until the old version is explicitly pruned. This
allows the old version to be in effect for the
duration of the migration process, and even afterwards,
to facilitate A/B testing, gradual feature rollout and
fast roll-back in case a bug is found in the new version.

The second change is a client-side change, because
restricted tweets were implemented as a client-side only
feature. Canceling the feature is as simple as rolling
back the code to its original version, before the change
was originally introduced.

While rolling back the change could have been done in a
system not supporting~\MI{} as well, doing so would have
had undesired consequences. In traditional applications,
where access control is coupled with authorization, the
two happen in distinct times. Authorization happens when a
user decides to designate a tweet as
\emph{restricted}. This restriction is enforced by the
application's access control, when another user is
attempting to read that tweet.

When removing both the authorization and access-control
portions of this feature, old tweets that were already
marked restricted by their authors would be visible to all
users. This is obviously not desired. To protect against
this, a traditional application would have to retain the
access-control logic handling restricted tweets for as
long as restricted tweets still exist as part of the
application's state.

With~\MI{}, access-control logic is constant, as defined
by \ref{def:MIP}. The~\MIP{} respects the readers-sets
assigned to facts. As long as restricted tweets are
retained, the~\MIP{} will respect their readers-sets and
only present them to authorized users. The application
completed its part when it created these facts with the
correct readers-sets.

\subsection{%
{Regulatory Compliance}
}\label{sec:disc-regulation}

In \TweetMI, only two kinds of facts comprise its user
data. In both kinds, the key happens to be the ID of the
user who creates (and thus ``owns'') them. If \TweetMI{}
were subject to regulatory requirements, such as GDPR's
\emph{right to erasure}, its developers would have no
problem providing a feature that allows a user to erase
all his or her data.

Unfortunately, in real-life applications there may be tens
to hundreds of different kinds of facts,\FOOTNOTE{One can
  draw intuition from the number of tables an application
  requires in traditional applications based on relational
  database.} where it is not guaranteed that the
user owning the fact can be directly inferred from its
data. This makes it significantly harder to comply with
such regulation.

However, every fact is annotated with a
writers-set. This set defines who owns the
fact. Having this annotation on every fact allows \Ax{}
itself to comply with the regulation, without involving
the application.

For example, to comply with the right to erasure, we can
add a feature to \Ax{} that will allow a user to erase all
facts that are owned solely by them. This can be done by,
e.g., adding a secondary index, indexing events by the
user ID in their writers-set. Adding this
index is a one-time effort, both in terms of programming
and in terms of applying it to existing data. Then, all
applications loaded on \Ax{} can enjoy the feature of
finding and removing all facts belonging to a single user.

\section{Validity and Threats to Validity}

\MI{} can be characterized as disruptive innovation~\cite{Christensen:1997:IDW}.
As such it starts
with a conceptual idea (``\emph{Eureka!}") and faces feasibility concerns
(``\emph{it will never work!}"), performance-related concerns (``\emph{it
might work but it cannot be made to work efficiently!}"), and concerns
that are social in nature (``\emph{No one would want or be able to use
it!}").

This paper presents the concept of~\MI{} and concentrates on the
feasibility phase (is it possible to support~\MI?). As evidence we submit a
substantial software artifact, \Ax, that validates the feasibility claim~\cite{Shaw:2002:WMG}.  We
make no claims regarding performance (e.g., runtime or memory usage) due
to a lack of a base line to compare against, although our 
design choices (e.g., the use of Apache Storm) enable scalability in principle. Nor
do we make any claims about the usability of our system (e.g., will
\SaaS{} programmers be willing to use \Clojure?), although our experience implementing
\TweetMI{} over~\MI{} was positive. Of course, future
work can build upon our work to propose ways to improve performance
and usability, comparing future implementations to ours.


\Ax{} is written in \Clojure.  We selected \Clojure{}
thanks to its combination of expressiveness on the one
hand, and the richness of its ecosystem on the other hand.
The choice of \Clojure{} was also driven by the ease of
creating DSLs, along with access to the vast \Java{}
ecosystem, available through \Clojure{}'s \Java{}
interoperability.  \Ax{}'s code consists of 3442 lines of
\Clojure{} code, and 258 lines of \ClojureScript{} code
(\Ax{}'s client-side library). Its documentation is bundled
with its tests, and the examples in the documents run as
\Ax{}'s unit tests, assuring these examples are always up
to date.\IGNORENOTE{Links to \Ax{}'s documentation can be
found in footnotes spread around this paper, 
\textcolor{Turquoise}{with author names temporarily removed
to allow double-blind peer-review}.}


The development of \Ax{} was driven by the five concerns related to
information that we introduced as criteria for~\MI. Obviously, there are
other concerns that could have been raised.  However, this threat
to validity is mitigated by the fact that abstracting any of these concerns is a
challenge in and by itself and collectively these concerns provide a
nontrivial benchmark.


Part of our observations about the behavior of~\MI{} is based on our own
experience developing \TweetMI{}. Naturally, a real-life application
such as the equivalent of Twitter or Facebook would be more demanding,
and these observations should be substantiated in other applications
too.  However, \TweetMI{} merely serves here as a part of the
demonstration of the feasibility of~\MI{},
for which even one example is meaningful.

\ref{tab:latency} shows latency measurements we took with \TweetMI. We
measured the latency in milliseconds it takes a tweet to arrive to the
same user's timeline (first row), and to a follower's timeline (second
row). In the former case, fact events describing the tweet are
processed by a single link to create the result event, which is sent
to the client. In the latter case, fact events arrive at a rule which
produces derived facts. These facts then arrive at another rule, which
creates the result events.

\begin{table}[tb]
  \centering
  \begin{tabular}{c|c|c|c}
    Number of Links & Min & Max & Average \\
    \hline
    1               & 84  & 154 & 117.2   \\
    2               & 108 & 186 & 144.3
  \end{tabular}
  \caption{Latency measurement results measured in milliseconds, averaged over 20 runs}
  \label{tab:latency}
\end{table}

The \texttildelow{30} milliseconds difference between the two measurements provides
an indication of the latency of a single link, including processing
time, and the latencies of the event broker and the database. Assuming
links are comparable (data passes through the same event broker and is
stored on the same database), about 85 milliseconds in each case are
attributed to the global overhead, including time spent on the client
(measurements were made end-to-end on the client side), communication
latency and time spent at the gateway tier. 

\section{Related Work}\label{sec:related-work}

While there is a multitude of approaches to application
development with significant trade-offs between one
another, the closest work to ours in terms of the problem
addressed is \Solid~\cite{Mansour:2016:DSP}.  \Solid{}
is a set of conventions that allow building social
applications in a decentralized manner.  But even \Solid,
which seems to address a similar need, actually adopts
a different approach. Specifically, \Solid{} allows each
user to choose his or her own \emph{information solution},
but this comes at the cost of not being able to run any
update-time logic on the server-side, which is a crucial
factor in large-scale applications.

In \Solid, the application is implemented on
clients. Users of \Solid{} applications can choose a
personal data store, which implements the \emph{Linked
Data Platform (LDP)} web standard to store their personal
data\URLNOTE{.}{https://github.com/linkeddata/cimba}
The application sends this server requests to update the
user's data, and queries, which are answered by following
links on the data, which may cross over to data stored on
other servers.

\Solid{} provides a data ownership model similar to our
own, in which the user is the sole owner of his or her
data, capable of restricting access and removing the data
at all times. In \Solid{} this is achieved by decoupling
the data storage concern from the application completely,
in the sense that users are free to choose their own
personal data stores, regardless of the applications
they use.

However, this freedom comes at a cost. With data for
different users residing on different servers, it is
unclear how efficient would a query in \Solid{} be, if
it spans different servers. Even if LDPs employ clever
heuristics to pre-fetch data from other LDPs if it seems
relevant to queries they are asked, it is still unclear
how data ownership is handled in such cases.

Furthermore, it is unclear how features that require
applying logic on data are implemented in applications
based on \Solid. For example, how can a \Solid{}
Twitter-like application index tweets according to users
mentioned in them. A
\Solid{} application can perform this denormalization in
a similar way to traditional applications, as tweets are
created, with similar limitations. However, it is unclear
how this can be done on existing data, when such a feature
is added as an evolutionary step when tweets already exist.

Similar to \Solid, an \Ax{} application is implemented
primarily on the client side. However, \Ax{} does
allow limited server-side logic to be specified by an
application. This logic allows data denormalization that
is missing in \Solid. This comes at the price that all
users using one application must trust the same \MIP{}
provider. However, this is still a big improvement relative
to the state of the art, in which users need to trust
the application provider for every application they use,
with their data.

\subsection{Triggers, Stored Procedures, and Row-Level
Security}

Relational databases have the ability to store data-aware
functions and update procedures, and to invoke them
automatically upon changes to the data. Some databases
go beyond this and support row-level security, where
user-provided functions are used to determine if a certain
user should or should not be allowed to access a certain
record.  These database features allow for developers
to make access control and denormalization part of the
database schema, thus freeing the logic tier from having
to worry about these aspects. In fact, such databases can
be accessed directly from the client-side, as they apply
access control by themselves.

However, compared to the traditional three-tier
architecture, in which access control and denormalization
are applied at the logic-tier, this is merely a change of
venue. The same logic, written in a different language,
runs in the data-tier instead of the logic-tier.

In comparison, \MI{} provides a separation between an
application-agnostic \MIP{} and application-specific
logic that does not have access to the data. Using stored
procedures and triggers may raise a database to the level
of an \AP{} (\ref{def:AP}, \ref{sec:ap}), but it needs
more than that to become a \MIP. Specifically, stored
procedures have unrestricted access to the database. Code
injection into a stored procedure (e.g., through the use
of Dynamic SQL) can provide an attacker the ability to
steal and corrupt user data.

\subsection{Access Control Lists}

With \MI{}, every piece of data is annotated with
a readers-set and a writers-set to specify user
permissions over it. This may seem similar to \emph{Access
Control Lists} (\emph{ACL})~\cite{Karjoth:2008:IAB},
an access-control approach which attaches a list of
allowed users for each operation (e.g., read, write,
manage permissions, etc) to each resource (e.g., file,
directory, record, etc).

While both approaches are similar in that permissions are
provided as annotations on objects, making access control
generic, these two approaches feature some fundamental
differences.

First, they are different in how they treat
integrity. \MI{} allows any user to create any piece of
data, given that the user attributes this piece of data
to herself. In contrast, ACL typically uses hierarchy to
authorize the creation of new objects. For example, the
ACL on a directory determines who is allowed to create
new files in it.

Second, ACL, does not have a notion of user
groups.
its group-aware variant --
ACLg, does have a notion of user groups, but
requires groups to be managed explicitly.
\MI, in contrast, allows for user group membership to be derived
automatically from the application logic.

Third, ACL does not have a sense of derived data, and does
not propagate permissions to computational products. In
\ref{sec:event-processing} we describe how the \MIP{}
propagates permissions to derived facts, so that the
application's implementation does not get a chance to get
it wrong.

\subsection{Access Control for SaaS}

Like ACL, some access control models have
been developed specifically for SaaS. S-RBAC is one,
which extends RBAC~\cite{Ferraiolo:1995:RBA} with a hierarchical model to address
applications in which multiple customers share the same
instance~\cite{Li:RBA:2010}.

Unfortunately, this model assumes that SaaS applications
maintain some of the properties of enterprise applications,
in that they serve organizations that have little
interaction with one another. Social networks are a
counter-example, since in a social network every single
user can be associated (follow, be friends with, etc.) with
any other user. As result, roles should be applied at the
granularity of a single user. At this level of granularity,
RBAC-base approaches lose their advantage relative to ACL.

\section{Conclusion}\label{sec:conclusion}

In this paper we present a novel way to abstract away
what is currently a core part of every application: the
\emph{management of information}. By doing so, we move many
of the concerns every application needs to deal with today,
to a single piece of software -- namely, the \MIP, similar
to how \MM{} moved memory safety concerns previously dealt
with by individual programs to a single piece of software
-- namely, the virtual machine.

Trust is a key concept in \MI{}. By not trusting
applications, the \MIP{} relieves the application of
its traditional responsibility as the gate keeper of
user data. A client representing a user is expected
to do so truthfully (i.e., create facts with readers
and writers-sets that correspond to the user's
expectations), but once the fact has been created,
it is no longer the application's concern.  This means
that the application has almost no consequence for any
of the various aspects of information security.  Indeed,
out of the nine application-level vulnerabilities listed
in the 2017's \emph{OWASP Top 10}~\cite{OWASP:2017:T10}
(\ref{sec:disc-vulnerabilities}), only one remains a
concern for \Ax{} applications, being a client-side
vulnerability.

Obviously, trust has to go somewhere, and in \MI{}
both users and application developers need to trust the
\MIP{}. However, the \MIP{} has more potential in becoming
trustworthy than any given application, since it has a
small attack surface relative to typical SaaS applications.

From the application developer's perspective, \MI{}
provides a way to enjoy the benefits of SaaS while
deferring much of its challenges to the \MIP{}. In many
ways, \MI{} brings software developers back to being
just that. They are entrusted to provide good software,
and are exempt from the need to host it, secure it, and
comply with data protection regulations.

From the end user's perspective, \MI{} brings an ownership
model in which users are once again in control over their
own data.



\def\dupcite#1{\textcolor{red}{Use instead BibTeX /cite\{\url{#1}\} \cite{#1}}}
  \def\usefootnote{\textcolor{red}{Use footnote instead}}
  \def\noopsort#1#2{#2}\wlog{aop.bib}\def\TODO#1{\framebox[1\columnwidth]{\parbox{.95\columnwidth}{\textcolor{red}{\small\textsf{TODO:
  #1}}}}}\wlog{self.bib}\wlog{misc.bib}\wlog{aopdebug.bib}\wlog{aop-dsal.bib}\wlog{aop-bibliography.bib}\wlog{acp4is.bib}\wlog{aosd-book.bib}

\nocite{Yang:2012:LAE}
\appendix
\section{\TweetMI{} Example}\label{appendix:tweetmi}

To give the reader a concrete sense of what it means to develop an
application under \MI{}, we describe here the implementation of
\TweetMI{}.
\ReF{appendix:spec} provides the complete requirements for this application.
\ReF{appendix:managing-state} describes how the client manages
state. \ref{appendix:logic} describes \TweetMI{}'s the update-time rules,
and \ref{appendix:queries} describes \TweetMI{}'s timeline query, and
the clauses that implement it.

\subsection{\TweetMI{} Specification}\label{appendix:spec}

\TweetMI's user interface has three panes: {\ONE}~a \emph{tweet pane}, in
which users can create, update or delete their own tweets, {\TWO}~a
\emph{following pane}, in which a user can control who they follow and
see who follows them, and {\THREE}~a \emph{timeline pane}, in which users
can see their timelines.

The timeline of user $A$ consists of tweets made by either {\ONE}~user
$B$, whom user $A$ follows, or {\TWO}~user $A$ him/herself. The timeline
is sorted according to the time of the tweet in descending order, so
that the most recent tweets will be presented first. Since the
complete timeline can be very big and contain many old tweets, \TweetMI{}
only shows the most recent tweets (initially, tweets from the last 7
days). If a user wishes to see older tweets, clicking a button at the
bottom of the list fetches older tweets.

The pane for user~$A$ displays two lists: {\ONE}~a list of
\emph{followers} (any user~$B$ who follows user~$A$), and a list of
\emph{followees} (any user~$C$ who user~$A$ follows). Next to each
user name there is a ``follow'' or an ``unfollow'' button, depending
on whether or not user~$A$ follows them already. Clicking a ``follow''
button attached to user~$B$ will cause user~$A$ to follow user~$B$,
causing user~$B$'s tweets to appear in user~$A$'s timeline, user~$B$
to appear in user~$A$'s followees list, and the button to change its
caption to ``unfollow''. Pressing an ``unfollow'' button attached to
user~$B$ causes user~$A$ to no longer follow user~$B$ and user~$B$'s
tweets to be removed from~$A$'s timeline.

Tweets can be either \emph{public}, i.e., visible to all users, or
\emph{restricted}, in which case they are only visible to
followees. For example, if user~$A$ follows user~$B$, and user~$B$
follows user~$C$, if user~$B$ tweets a public tweet, user~$A$ will see
it. However, if user~$B$ tweets a restricted tweet, user~$A$ will not
be able to see it. Only if user~$C$ starts following user~$B$ can 
user~$C$ see user~$B$'s restricted tweets.\footnote{Twitter uses a similar
criterion for authorizing private
messages: {\url{https://help.twitter.com/en/using-twitter/direct-messages}}}

\subsection{Managing State}\label{appendix:managing-state}

In an \MI{}
application, the client is responsible for managing data for the user
it represents. In \TweetMI{}, the client is written in
\ClojureScript{}.
To synchronize the state with the user interface, \TweetMI{} uses
\Reagent\footnote{\url{https://github.com/reagent-project/reagent}} a
\ClojureScript{} adapter for
React.\footnote{\url{https://reactjs.org}} \React{} and \Reagent{} allow the user
interface to be defined as declarative functions, mapping the state
into its presentation, in the form of HTML elements. Behind the
scenes, these libraries are responsible for tracking changes in the
data, mapping them to changes in the presentation.

\subsubsection{Managing Tweets}
\paragraph{View Definition}
\ref{lst:appendix:tweets-view} shows the view definition used by the
\TweetMI{} client to bind tweets made by a certain user. \ReF{TV:01} defines
the view's name and arguments (the argument \lstinline$me$ in this
case, representing the user ID).

\begin{figure}
  \def\lstcaption{The \lstinline$tweets$ view definition}
  \centering
  \begin{spacing}{\spc}
\begin{lstlisting}[caption=\lstcaption,label=lst:appendix:tweets-view]
(defview tweet-view [me]				^\label{TV:01}^
  [:tweetmi/tweeted me text ts attrs]			^\label{TV:02}^
  :store-in (reagent/atom nil)				^\label{TV:03}^
  :order-by (- ts)					^\label{TV:04}^
  :writers #{$user}					^\label{TV:05}^
  :readers (if (:restricted attrs)			^\label{TV:06}^
             #{[:tweetmi.core/follower me]}		^\label{TV:07}^
             #{}))					^\label{TV:08}^
\end{lstlisting}
  \end{spacing}
\end{figure}

\ReF{TV:02} provides the pattern of the
fact this view corresponds to. In this case, the fact's name is
\lstinline$tweetmi:tweeted$, its key (the user who wrote the tweet) is
bound to the view's parameter, \lstinline$me$. This means that the
view represents, at one time, only tweets by a single user -- the user
logged in to the client. The view's data arguments, \lstinline$text$,
\lstinline$ts$, and \lstinline$attrs$ are given as well.

\ReF{TV:03,TV:04,TV:05,TV:06,TV:07,TV:08} define the view's
optional properties. \ReF{TV:03} assigns a Reagent atom as the mutable
container for the view. This allows Reagent to track updates to the
state, and reflect them to the presentation.

\ReF{TV:04} defines the sorting order. By sorting according to
\lstinline$(- ts)$ (the negated timestamp value) we actually sort the
facts in descending time order.

\ReF{TV:05,TV:06,TV:07,TV:08} define the readers and writers set for the
tweet.
On \ref{TV:05} we attribute the tweet to the
logged-in user (represented by \lstinline!$user!
), by setting the writer set to be a user group that consists of only
that user. This value is the default, and is given here for
introductory purposes only.

\ReF{TV:06,TV:07,TV:08} define the reader set conditionally,
depending on the optional attribute \lstinline$:restricted$. If this
attribute is present and \lstinline$true$, the reader set is set to a
parametric group \mbox{\lstinline$[:tweetmi.core/follower me]$} consisting of users who
the current user (\lstinline$me$)
follows.
If the attribute
\lstinline$:restricted$ is false or is not defined, we define the
readers set to be the universal set~\lstinline$#{}$. An
interset is a set represented as an intersection of groups. If no
groups are intersected, the set is universal.

\paragraph{Using the View Function}
The view is represented as a \ClojureScript{} function. This function,
which derives its name from the view, takes $n+1$ arguments, where $n$
is the number of parameters the view takes. As first argument, the
function takes a \lstinline$host$ map, which represents the connection
to \Ax{}. The rest of the arguments correspond to the view
parameters. The view function returns a sequence of records (\Clojure{}
maps) representing the different facts in the view.

\ref{lst:appendix:tweets-ui} shows the function that renders the user
interface that allows users to view and edit their own tweets. The
function takes as parameter the \lstinline$host$ map, and returns a
nested structure of \Clojure{} vectors representing an HTML
fragment. \Reagent, along with the underlying
\React, is responsible to call this
function for every state change that affects it, and to translate the
vectors into DOM\footnote{Document Object Model (DOM): A tree of
  visual objects that correspond to elements in an HTML document,
  facilitating programmatic manipulation.} updates.

\begin{figure}
  \def\lstcaption{Component function for editing tweets}
  \centering
  \begin{spacing}{\spc}
\begin{lstlisting}[caption=\lstcaption,label=lst:appendix:tweets-ui]
(defn tweets-pane [host]							^\label{TU:01}^
  (let [tweets (tweet-view host (user host))					^\label{TU:02}^
        {:keys [add]} (meta tweets)]						^\label{TU:03}^
    [:div									^\label{TU:04}^
     [:h2 "Tweets"]								^\label{TU:05}^
     [:button {:on-click #(add {:text ""					^\label{TU:06}^
                                :ts ((:time host))				^\label{TU:07}^
                                :attrs {}})} "tweet!"]				^\label{TU:08}^
     [:button {:on-click #(add {:text ""					^\label{TU:09}^
                                :ts ((:time host))1				^\label{TU:10}^
                                :attrs {:restricted true}})}			^\label{TU:11}^
      "tweet restricted!"]							^\label{TU:12}^
     [:ul									^\label{TU:13}^
      (for [{:keys [me text ts attrs swap! del!]} tweets]			^\label{TU:14}^
        [:li {:key ts}								^\label{TU:15}^
         [:input {:value text							^\label{TU:16}^
                  :on-change #(swap! assoc :text (.-target.value %))		^\label{TU:17}^
                  :style {:color (if (:restricted attrs)			^\label{TU:18}^
                                   "red" "black")}}]				^\label{TU:19}^
         [:button {:on-click del!} "X"]])]]))					^\label{TU:20}^
\end{lstlisting}
  \end{spacing}
\end{figure}

\ReF{TU:02} calls the view function \lstinline$tweet-view$, giving it the
\lstinline$host$ map and the current user \mbox{\lstinline$(user host)$} as
parameters. The sequence of records is stored in variable
\lstinline$tweets$. \ReF{TU:03} extracts the function \lstinline$add$ from
the returned sequence's metadata. This function adds a new record to
the view. It is used in \ref{TU:06,TU:09} in the \lstinline$:on-click$ callback
functions associated with the \lstinline$"tweet!"$ and
\mbox{\lstinline$"tweet restricted!"$} buttons. When either button is
clicked, a new record with an empty \lstinline$text$ field and a
current time (returned by the \lstinline$:time$ function provided by
the \lstinline$host$ map) is created. The \mbox{\lstinline$"tweet restricted!"$}
button also sets the \lstinline$:restricted$ attribute
to \lstinline$true$.

\ReF{TU:13,TU:14,TU:15,TU:16,TU:17,TU:18,TU:19,TU:20} display the list of existing tweets, allowing the
user to update and delete each of them. \linereF{lst:appendix:tweets-ui}{TU:14} uses \Clojure{}'s
\lstinline$for$ macro, binding the \lstinline$text$, \lstinline$ts$
and \lstinline$attrs$ fields from each record to variables of the same
names. In addition, two functions: \lstinline$swap!$ and
\lstinline$del!$ which are provided by the view are also extracted. On
\ref{TU:15} a list item (\lstinline$:li$) is rendered. \Reagent{} requires
that each item in a list (as the one returned by the \lstinline$for$
macro) have a unique key, so we use \lstinline$ts$ for this
purpose. \ReF{TU:16,TU:17,TU:18,TU:19} define an input box for editing the
tweet. Its \lstinline$:value$ attribute is set to the \lstinline$text$
field, to show the content of the tweet. Its \lstinline$:on-change$
event handler uses the record's \lstinline$swap!$ function to update
this record, updating its \lstinline$:text$ field to the
\lstinline$.value$ attribute of the input box at the time the event
fired. Calling the \lstinline$swap!$ function causes the associated
fact to be updated both locally on the client, and in \Ax{}, by
generating a change event and sending it to the host. The
\lstinline$:style$ attribute defined on \andref{lst:appendix:tweets-ui}{TU:18}{TU:19} sets the
tweet's text color to red or black, depending on whether the tweet is
restricted or not.
\ReF{TU:20} creates a button associated with the \lstinline$del!$
function, which when invoked deletes the associated fact.

\subsubsection{Follow Button}
\ref{lst:appendix:follow-button} shows the function
rendering the follow button, and its associated view. The \lstinline$follow-view$ view defined in 
\rangeref{lst:appendix:follow-button}{FB:01}{FB:03} defines a view for \lstinline$:tweetmi/follows$
facts, that indicate that user \lstinline$u1$ follows user
\lstinline$u2$. Unlike the \lstinline$tweet-view$ view defined in
\ref{lst:appendix:tweets-view}, here the view arguments appear in both the key
(\lstinline$u1$) and the data elements (\lstinline$u2$) of the fact
pattern. This means that the view is Boolean. It indicates whether
or not \lstinline$u1$ follows \lstinline$u2$. If yes, the sequence
returned by the view function will contain one element, and if not,
the sequence will be empty.

\begin{figure}
  \def\lstcaption{Follow button}
  \centering
  \begin{spacing}{\spc}
\begin{lstlisting}[caption=\lstcaption,label=lst:appendix:follow-button]
(defview follow-view [u1 u2]				^\label{FB:01}^
  [:tweetmi/follows u1 u2]				^\label{FB:02}^
  :store-in (reagent/atom nil))				^\label{FB:03}^
							^\label{FB:04}^
(defn follow-button [host u1 u2]			^\label{FB:05}^
  (let [f (follow-view host u1 u2)			^\label{FB:06}^
        {:keys [add]} (meta f)]				^\label{FB:07}^
    (cond (> (count f) 0)				^\label{FB:08}^
          (for [{:keys [del!]} f]			^\label{FB:09}^
            [:button {:key 0				^\label{FB:10}^
                      :on-click del!} "unfollow"])	^\label{FB:11}^
          :else						^\label{FB:12}^
          [:button {:on-click #(add {})} "follow"])))	^\label{FB:13}^
\end{lstlisting}
  \end{spacing}
\end{figure}

\ReF{FB:05,FB:06,FB:07,FB:08,FB:09,FB:10,FB:11,FB:12,FB:13}
show the component function that renders the follow/unfollow button.
It takes as arguments both the
current user (\lstinline$u1$) and the user to follow or unfollow
(\lstinline$u2$), and renders, based on whether the sequence is empty
or not a ``follow'' button, which when clicked will add an element to
the collection, effectively making \lstinline$u1$ follow
\lstinline$u2$, or an ``unfollow'' button, which when clicked will
delete the single element that exists in the sequence, making
\lstinline$u1$ stop following \lstinline$u2$.

\subsection{Update-Time Logic}\label{appendix:logic}
Update-time logic is required to perform computation for which waiting
to query-time will be too late. In the case of \TweetMI{}, this
includes pre-calculating user timelines, based on users they follow
and their tweets, and indexing tweets according to text that appears
in them.

The complete logic needs to take care of one
additional concern: \emph{pagination}.

\paragraph{Pagination}
Pagination is the practice of limiting the amount of data received
from a server, to save resources on both the client and the server. A
practical application needs to make sure that any given client request
is answered with a bound amount of data. This can be addressed as a
generic server-side rule, limiting all responses to a pre-determined
size, but by doing so we may be leaving out important information.

To fix this problem for \TweetMI{}, we index tweets according to the
day in which they were created. Each tweet has a \lstinline$ts$ field,
which contains the UNIX time (in milliseconds) of its creation. By
dividing this number by the number of milliseconds in a day (integer
division) we can get a number that can act as an effective key for
tweets. Users are often interested in recent tweets and are less
interested in older ones. A timeline is therefore organized in
time-descending order. To only get tweets we need we can only fetch
tweets that were created on a certain day-range. We can extend that
range as users look deeper and deeper into their timelines.

\subsubsection{Followee Tweets}\label{appendix:joins}
\ref{lst:appendix:followee-tweet-rule} shows the definition of the
\lstinline$followee-tweets$ rule. \ReF{FT:01,FT:02} define the helper
function \lstinline$ts-to-day$, which converts a timestamp into a day
number, as needed for pagination. This is an ordinary \Clojure{} function.
The rule itself is given on \rangeref{lst:appendix:followee-tweet-rule}{FT:04}{FT:07}. \ReF{FT:04} defines
the format of the derived facts to be generated by this rule. The key
to these facts is compound, consisting of both the \lstinline$user$ ID
for the user whose timeline this is, and the \lstinline$day$ number in
which the tweet has been written. The data elements are the
\lstinline$author$ of the tweet, its \lstinline$text$, and its
timestamp (\lstinline$ts$).

\begin{figure}
  \def\lstcaption{The followee-tweet rule}
  \centering
  \begin{spacing}{\spc}
\begin{lstlisting}[caption=\lstcaption,label=lst:appendix:followee-tweet-rule]
(defn ts-to-day [ts]						^\label{FT:01}^
  (quot ts msec-in-day))					^\label{FT:02}^
								^\label{FT:03}^
(clg/defrule followee-tweets [[user day] author text ts]	^\label{FT:04}^
  [:tweetmi/follows user author] (clg/by user)			^\label{FT:05}^
  [:tweetmi/tweeted author text ts attrs] (clg/by author)	^\label{FT:06}^
  (let [day (ts-to-day ts)]))					^\label{FT:07}^
\end{lstlisting}
  \end{spacing}
\end{figure}

On \ref{FT:04} we use a \lstinline$let$ guard
to calculate the
tweet's \lstinline$day$ based on its timestamp, by calling
\lstinline$ts-to-day$.

\subsection{Queries and Clauses}\label{appendix:queries}
Now that we have defined our rules for creating derived facts, we can
define the timeline query to aggregate them.

\ref{lst:appendix:timeline-query} shows the query definition for the user
timeline in \TweetMI. \ReF{FB:01} defines the query's name and parameters,
including the user ID (\lstinline$me$) and two parameters to identify
the time-range we are interested in. These parameters
(\lstinline$day-from$ and \lstinline$day-to$) are used for pagination,
to limit the
query results to only those of the last few days.
A user can request older tweets by clicking
a button at the bottom of the list.

\begin{figure}
  \def\lstcaption{Timeline query}
  \centering
  \begin{spacing}{\spc}
\begin{lstlisting}[caption=\lstcaption,label=lst:appendix:timeline-query]
(defquery timline-query [me day-from day-to]			^\label{TQ:01}^
  [:tweetmi/timeline me day-from day-to -> author tweet ts]	^\label{TQ:02}^
  :store-in (reagent/atom nil)					^\label{TQ:03}^
  :order-by (- ts))						^\label{TQ:04}^
\end{lstlisting}
  \end{spacing}
\end{figure}

\ReF{FB:02} provides the query's pattern, taking the user and the day
range as input, and the tweet details as output.
Similar to the view definition in \ref{lst:appendix:tweets-view},
\ref{FB:03,FB:04}
define the storage to use a \Reagent{} atom, and the order
to be descending by time.

Like a view, a query defines a function which can be called by the
functions rendering the user interface. Unlike a view, the records
returned by a query do not contain mutation functions
(\lstinline$swap!$ and \lstinline$del!$), since the
results cannot be updated by the client.

\subsubsection{Clauses}\label{appendix:clauses}
\begin{figure}
  \def\lstcaption{Timeline clause}
  \centering
  \begin{spacing}{\spc}
\begin{lstlisting}[caption=\lstcaption,label=lst:appendix:timeline-clause]
(defn days-in-range [from-day to-day]				^\label{TC:01}^
  (let [day-range (range from-day to-day)]			^\label{TC:02}^
    (if (> (count day-range) 20)				^\label{TC:03}^
      []							^\label{TC:04}^
      day-range)))						^\label{TC:05}^
								^\label{TC:06}^
(clg/defclause tl-1						^\label{TC:07}^
  [:tweetmi/timeline user from-day to-day -> author text ts]	^\label{TC:08}^
  (for [day (days-in-range from-day to-day)])			^\label{TC:09}^
  [followee-tweets [user day] author text ts])			^\label{TC:10}^
								^\label{TC:11}^
(clg/defclause tl-2						^\label{TC:12}^
  [:tweetmi/timeline user from-day to-day -> user text ts]	^\label{TC:13}^
  [:tweetmi/tweeted user text ts attrs] (clg/by user)		^\label{TC:14}^
  (for [day (days-in-range from-day to-day)])			^\label{TC:15}^
  (when (= (ts-to-day ts) day)))				^\label{TC:16}^
\end{lstlisting}
  \end{spacing}
\end{figure}

In the server-side logic, clauses 
contribute answers to queries.
\ref{lst:appendix:timeline-clause} shows two clauses that contribute
results to the \lstinline$:tweetmi/timeline$ query.
\rangeref{lst:appendix:timeline-clause}{TC:01}{TC:05} define a function
\lstinline$days-in-range$, which converts the boundaries
\lstinline$from-day$ and \lstinline$to-day$ to a \Clojure{} sequence of
days in that range. It checks that the range does not span over 20
days to protect against abuse.

The clause defined on \rangeref{lst:appendix:timeline-clause}{TC:07}{TC:10} contributes results based on
the \lstinline$followee-tweets$ rule defined in
\ref{lst:appendix:followee-tweet-rule}. \ReF{TC:07} provides the clause's name,
which is not referenced anywhere. \ReF{TC:08} provides the signature of the
query this clause is intended to answer. This signature matches the
one defined in the query definition in
\ref{lst:appendix:timeline-query}. \ReF{TC:09} uses a \lstinline$for$ guard to
iterate over the different days in the range, which is calculated by
calling \lstinline$days-in-range$. \ReF{TC:10} references the
\lstinline$followee-tweets$ rule, providing the user ID and a concrete
day as key.

The clause defined on \rangeref{lst:appendix:timeline-clause}{TC:12}{TC:16} contributes to the timeline
tweets made by the user whose timeline we are querying. It makes
direct access to \lstinline$:tweetmi/tweeted$ facts, guarded by a
\lstinline$by$ guard to make sure the tweets were indeed made by the
user in question. Then the \lstinline$for$ guard and the
\lstinline$when$ guard are used together to make sure the day that
corresponds to the tweet's timestamp (\lstinline$ts$) is within the
requested day range.

\end{document}